\documentclass[12pt,preprint]{aastex}

\usepackage{epsf}

\def\etal{{\it et al.}}
\def\eg{{\it e.g.,}}
\def\ie{{\it i.e.,~}}
\def\kms{~{\rm km~s^{-1}}}
\def\K{~{\rm K}}

\begin{document}

\title{Shock-Heated Gas in the Large Scale Structure of the Universe}

\author{Hyesung Kang\altaffilmark{1},
        Dongsu Ryu\altaffilmark{2},
        Renyue Cen\altaffilmark{3},
    and Doojong Song\altaffilmark{4}}
\altaffiltext{1}
{Department of Earth Sciences, Pusan National University, Pusan 609-735,
Korea:\\ kang@uju.es.pusan.ac.kr} 
\altaffiltext{2}
{Department of Astronomy \& Space Science, Chungnam National University,
Daejeon 305-764, Korea:\\ ryu@canopus.chungnam.ac.kr}
\altaffiltext{3}
{Princeton University Observatory, Princeton, NJ 08544:
cen@astro.princeton.edu}
\altaffiltext{4}
{Korea Astronomy Observatory, Taejeon 305-348, Korea: djsong@kao.re.kr}

\begin{abstract}
Cosmological shock waves are a ubiquitous consequence of cosmic structure
formation. They play a major role in heating baryonic gas in the large
scale structure of the universe. In this contribution we study the
shock-heated gas in connection with shocks themselves, using a set of
N-body/hydrodynamic simulation data of a $\Lambda$CDM universe. The
distributions of shock speed and temperature of shock-hated gas both
should reflect the depth of gravitational potential of associated
nonlinear structures, so their morphology. From their distributions we
find, in addition to hot gas in and around clusters/groups and warm-hot
intergalactic medium (WHIM) with $T=10^5-10^7 \K$ mostly in filaments,
there is a significant amount of low temperature WHIM with $T < 10^5\K$
distributed mostly as sheet-like structures. The low temperature WHIM
was heated and collisionally ionized mainly by shocks with
$v_{sh} \la 150 \kms$, while photo-ionization by the UV and X-ray
background radiations is important for metal ions. Not only the WHIM with
$T=10^5-10^7 \K$ but also the WHIM with $T < 10^5\K$ make up a significant 
fraction of gas mass, implying the low temperature WHIM could be important
in resolving the missing baryon problem. The shock-heated gas
in filaments and sheets are manifested best through emissions and
absorptions in soft X-ray and far UV. We confirm that the WHIM with
$T=10^5-10^7 \K$ makes significant contributions to the soft X-ray
background, absorptions of highly ionized species such as OVII and OVIII
in AGN spectra, and line emissions from OVII and OVIII ions, as pointed
by previous studies. However, the WHIM with $T < 10^5\K$ is the major
contributor to absorptions of lower ionized species such as OV and OVI,
because these photo-ionized ions are most abundant in sheets of low
density and temperature. On the contrary, lines of OV (630 \AA) and
OVI (1032 \AA) are emitted mostly from the WHIM with $10^5 < T < 10^6\K$,
because they are from collisionally excited ions.

\end{abstract}

\keywords{cosmology: theory -- diffuse radiation -- large-scale structure
of universe -- methods:numerical -- quasars: absorption lines}

\section{Introduction}

In the current paradigm of the cold dark matter cosmology, the large
scale structure of the universe has formed through hierarchical
clustering of matter. Deepening of gravitational potential wells
caused baryonic matter to accrete onto nonlinear structures. It was
postulated that the baryonic matter behaved like a gas and
``collisionless shocks'' formed in the tenuous, magnetized, cosmic
plasma via collective electromagnetic viscosities
\citep{krj96,qis98,mrk00,gp03,rkhj03}. In fact, the gravitational
energy of the baryonic gas associated with structure formation was
dissipated into heat primarily via such shocks. The dissipated energy,
then, governs the thermal history and nature of the gas in the universe,
and manifests the large scale structure as well as its dynamics through
emitted radiation \citep{kcor94,ckor95,co99a,dcob01,vss02,fl03}.

In a recent work, \citet{rkhj03} studied the properties of
``cosmological shock waves'' and their roles on thermalization of gas
and acceleration of nonthermal particles in the large scale structure, 
using cosmological N-body/hydrodynamic simulations. The cosmological
shocks were classified into two types. ``External shocks'' form around
outermost surfaces that encompass nonlinear structures, so they are
by nature accretion shocks that decelerate the never-shocked
intergalactic gas infalling toward sheets, filaments and knots.
``Internal shocks'' are produced within those nonlinear structures by
infall of previously shocked gas to filaments and knots, during
subclump mergers, as well as by chaotic flow motions. External shocks
have high Mach numbers of up to $M \sim 100$ due to low temperature
of the accreting gas. Internal shocks, on the other hand, have mainly
low Mach numbers of $M \sim$ a few, because the gas inside nonlinear
structures has been previously heated by shocks (external and/or
internal) and so has high temperature. However, internal shocks are
more important in energetics, because of higher preshock density.
As a result, thermalization of gas and acceleration of cosmic rays
occurred mostly at internal shocks.

The intergalactic gas was heated mostly by cosmological shock waves.
The shock-heated gas with $T=10^5-10^7 \K$ was studied previously,
utilizing cosmological hydrodynamic simulations \citep{co99a,dcob01}.
This component, which is called a warm-hot intergalactic medium (WHIM),
was estimated to contain $\sim 30 \%$ of baryon mass at the present
epoch and resides mostly in filaments with a median overdensity of
$\sim 10-30$. Such WHIM can count for a significant fraction of
the missing baryons at low redshifts \citep{fhp98,fuk03}. X-ray emission
from the WHIM may contribute to a fraction of the soft X-ray background
radiation (XBR) according to \citet{poc01}. While there were recent
reports of observations of the WHIM around clusters
\citep[see, \eg][]{fbh03,zmmg04b}, detection of such emission still
remains a challenging task due to low surface brightness. On the other
hand, the WHIM may be better detected through absorption systems and
emission lines in soft X-ray and far UV. For instance, the possibility
of probing the WHIM with OVI absorbers was explored by \citet{cen01}
and \citet{fb01}. Absorption systems due to other Oxygen and metal
ions in the intervening WHIM were studied by \citet{hgm98}, \citet{cwkd03}
and \citet{vbcm03}. Detections of such absorbers in X-ray observations
including Chandra were reported
\citep[see, \eg][among many]{nica03,mymg03}. Emission lines from
metal ions in the WHIM were studied by \citet{yyso03}, \citet{fssh03},
and \citet{fcsh03}. A possible detection of such emission lines from a
filament around Coma was reported \citep{fbh03}.

In this paper, we study the shock-heated gas in the
large scale structure of the universe and evaluate its observational
manifestations with data from a set of cosmological hydrodynamic
simulations where shock waves were identified \citep[see][]{rkhj03}.
Specifically, the properties of shock-heated gas are analyzed in
connection with those of cosmological shock waves. From the shock
speed and gas temperature distributions, we see that 1) the shocks
with $v_{sh} \ga 700 \kms$ and the hot gas with $T > 10^7\K$ are
distributed in and around clusters/groups, 2) the shocks with
$150 \la v_{sh} \la 700 \kms$ and the WHIM with $T=10^5-10^7 \K$
are mostly in filamentary structures, and 3) the shocks with
$v_{sh} \la 150 \kms$ and the WHIM with $T < 10^5\K$ are mostly in
sheet-like structures. Here we refer the uncollpased, collisionally
ionized, shock-heated gas as the WHIM and extend the low temperature
bound of the WHIM to well below $10^5$K. While the WHIM with
$T=10^5-10^7$ K has been studied extensively in many previous works,
the ``low temperature WHIM'' with $T<10^5$K has not received much
attention and will be the main subject of this study.

In the next section, simulation data are described along with numerical
details. The properties of shock-heated gas are described in \S 3,
and its observational manifestations are discussed in \S 4. Summary is
followed in \S 5.

\section{Simulation Data}

We used the data from a set of N-body/hydrodynamic simulations of
a $\Lambda$CDM universe with radiative cooling. The simulations
were performed using a PM+Eulerian hydrodynamic cosmology code that
was specifically designed to capture shocks with a high accuracy. It is
an updated version of the code described in \citet{rokc93}. For instance,
the code now adopts the MC (monotonized central difference) limiter,
instead of the original minmod limiter to achieve sharper resolution
of discontinuities \citep[see, e.g.,][]{leve97}. The $\Lambda$CDM model
adopted the currently favored values of the following cosmology parameters:
$\Omega_{BM}=0.043$, $\Omega_{DM}=0.227$, $\Omega_{\Lambda}=0.73$
($\Omega_{BM} + \Omega_{DM} + \Omega_{\Lambda}=1$),
$h \equiv H_0$/(100 km/s/Mpc) = 0.7, and $\sigma_8 = 0.8$.
These values are consistent with those fitted with the recent WMAP data
\citep[see, \eg][]{bhhj03,svpk03}. A cubic region of size
$L = 100 h^{-1}{\rm Mpc}$ at present was simulated inside the computational
box with $1024^3$, $512^3$ and $256^3$ grid zones for gas and gravity
and with $512^3$, $256^3$ and $128^3$ particles for dark matter,
allowing a ``fixed'' spatial resolution of $\Delta l = 97.7 h^{-1}$
kpc $-$ $390.6 h^{-1}$ kpc. We note that the simulations of different
resolution have the exactly same large scale structure. The only
difference is the Nyquist frequency in the initial condition, apart
from spatial and mass resolutions.

We adopted the tabulated cooling rate that had been calculated by
following the non-equilibrium collisional ionization of gas
with a given metallicity, cooling from $10^{8.5}$K to $10^4$ K under
the isobaric condition \citep{sd93}. The radiative cooling rate based
on equilibrium collisional ionization is not a good approximation
for $T \la 10^6 \K$, since recombination lags cooling and so ionization
fractions depend on thermal and ionization history of gas. The
non-equilibrium cooling rate for the gas cooling under the isobaric
condition, however, becomes a function of temperature only, if
the initial temperature is high enough to ensure the initial ionization
equilibrium (\eg~ $T \ga 10^6 \K$) and if only two-body collisional
processes are included \citep[see, \eg][]{sd02}. In our simulations,
gas was allowed to cool only down to $T \sim 10^4 \K$ and cooling was
set equal to zero for $T<10^4$K. For void gas, the minimum temperature
was set as the temperature of cosmic microwave background radiation,
\ie $T_{\rm min}=T_{CMB}(z)$. For metallicity, we adopted the mean
metallicity-density relation,
\begin{equation} 
{Z(\rho_{\rm gas}) \over Z_{\sun}} = \min\left[0.3, \max\left[0.006, 0.03
\left({\rho_{\rm gas} \over \bar \rho_{\rm gas}}\right)^{1/3}\right]\right],
\label{zmetal}
\end{equation}
which had been suggested by \citet{co99b}, using the data from
a cosmological simulation with feedbacks into the IGM from stars. 
The simulations are the same ones previously used in \citet{rkhj03} for
the study of cosmological shock waves, except radiative cooling
is included.

A few points on numerical details are noted.
1) Our simulations did not include star and galaxy formations and
feedbacks from those stars, galaxies and AGNs, nor photo-ionization
heating of the intergalactic gas. This is because our primary focus is
the heating by cosmological shock waves and the properties of
shock-heated gas in filaments and sheets. Non-gravitational feedbacks
are expected to be most important in high density regions inside clusters
and groups. Through previous studies by \citet{co99a} and \citet{dcob01},
it is known that the intergalactic gas was heated primarily by cosmological
shock waves. Feedbacks may have contributed only $\sim 20\%$ or so of the
thermal energy of the intergalactic gas, so they should have played only
minor roles in the thermal evolution of the WHIM in filaments
and sheets. On the other hand, photo-ionization heating should have
affected mostly diffuse gas in void regions and Ly $\alpha$ clouds, 
heating it to $T \sim 10^4 \K$ \citep[see, \eg][]{vss02}. But it should
not have affected much the properties of the WHIM, since most of
the WHIM has $T \ga 10^4 \K$ in our simulations. When we estimated
the amount of shock-heated gas, we excluded the components that would
have been heated primarily by photo-ionization (see the next section
for details). However, the photo-ionization of metal ions, such as
Oxygen, due to the background radiation was taken into account in
post-processing data analyses, since such photo-ionization dominates
over collisional ionization for the WHIM in filaments and sheets.
2) Because of the fixed-grid-based nature, our data have a limitation
in studies of high density regions such as cores of clusters and groups
of galaxies, compared to those using SPH or AMR codes \citep{kocr94,fwbb99}. 
However, the same nature of fixed grid-spacing turns out to be an
advantage in studying the low density gas in filaments and sheets. 
The typical thickness of filaments and sheets is $\sim 1 h^{-1}{\rm Mpc}$
or larger (see Fig. 2 below). In addition, the internal shocks inside
those nonlinear structures have a mean separation of
$\sim 1 h^{-1}{\rm Mpc}$ too \citep{rkhj03}. As shown below, numerical
resolution is important in studies of the gas inside those nonlinear
structures. The fixed grid-spacing of
$\Delta l = 97.7 h^{-1}$kpc, $\sim 1/10$ of those scales, in our highest
resolution simulation seems to be good enough to resolve filaments
and sheets and shocks in those structures \citep{rkhj03}.
3) In order to establish the connection between shock-heated gas and 
cosmological shocks, we identified shock surfaces by the procedure
described in \citet{rkhj03}. Only the shocks with $v_{sh} \geq 15 \kms$
were identified, although the code itself captures shocks of any speed.

\section{Shock-Heated Gas in Filaments and Sheets}

\subsection{Temperature and Density of Shock-Heated Gas}

We begin our discussion with the temperature and density of
shock-heated gas in the simulated universe. Figure 1 shows the mass
distribution in the plane of gas density and temperature at $z=0$ from
different simulations. The purple line along $T \sim 10^4$ K in high
density regime in the simulations with $1024^3$ and $512^3$ grid zones
is an artifact of numerical simulation. In order to isolate the
shock-heated WHIM, we draw two lines in the figure: 
1) The straight line represents the ``Lyman-$\alpha$ equation
of state'' of \citet{vss02},
\begin{equation}
T_{\alpha} = 0.418 \times 10^4 \left( \rho_{\rm gas} \over
{\bar \rho}_{\rm gas} \right)^{0.59},
\end{equation}
which was adjusted for different values of cosmological parameters
employed \citep[see also][among many references]{hg97}. The line
comes from the interplay of radiative heating by the UV background
with the expansion of the universe. The region of overdensity
$\delta \la 10$ that lies below this line would have been heated
primarily by the background UV radiation to the temperature depicted
by the line.
2) The curved line in the high density regime delineates the
``fast-cooling'' dense region, which was adopted also from \citet{vss02}
again with a suitable modification due to different values of parameters.
It comes from the equality of the effective cooling time due to radiative
cooling plus heating by the background UV radiation and the Hubble time.
The high density region ($\delta \ga 10$) below the Lyman-$\alpha$ line
and right to the fast-cooling line would have been collapsed into objects
such as stars and galaxies or would be in the form of bound objects,
if resolution were not limited by fixed grid-spacing in our simulations
\citep[see \eg][for the distribution of gas from an SPH simulation]{fssh03}.
Since our focus is on the intergalactic gas that was mainly heated
by shock waves, we exclude in our discussion and analyses below
the photo-heated gas as well as the gas in collapsed/bound objects.

Several points are apparent in Figure 1.
1) In addition to the hot component with $T > 10^7 \K$ and the WHIM
with $T=10^5-10^7 \K$, which were discussed in previous studies
of shock-heated gas \citep{co99a,dcob01}, there exists a significant
amount of low temperature WHIM in the range of $T < 10^5 \K$. The mass
fraction of the gas in each component is the followings: the hot
component $\sim 2.2 \%$, the WHIM with $T=10^5-10^7 \K$ $\sim 24 \%$
and the the WHIM with $T < 10^5 \K$ $\sim 13 \%$ in our highest
resolution simulation data (see Table 1). In the hot component, the
gas within 1 $h^{-1}$Mpc of clusters/groups hotter than 1keV was not
included. Most of the low temperature WHIM has the temperature of
$10^4 \K < T < 10^5 \K$, while the mass of the WHIM with
$T < 10^4 \K$ is $\sim 3 \%$ of total gas mass.
2) Our estimate of mass fraction for the WHIM with $T=10^5-10^7 \K$
is comparable to, although a little smaller than, $\sim 30 \%$ from
simulations using the PTreeSPH and AMR codes \citep{dcob01}. A part
of the difference would be due to feedbacks from stars, galaxies,
and AGNs, which were ignored in our simulations.
3) It is interesting to note that the distribution of our WHIM
follows very closely the ``equation of state for the warm IGM'',
analytically suggested by \citet{vss02}. However, their estimation of
mass fraction for the ``warm IGM'', $\sim 24 \%$, is somewhat smaller
than our estimation for all the WHIM with $T < 10^7 \K$.
4) Gas density reaches to higher values and cooling becomes even more
efficient with the $\rho^2$ dependence at higher resolution.
The resulting trend is that there is more gas with high density and
low temperature in higher resolution simulations. As a result,
the mass fraction of the WHIM continues to decrease as resolution
increases in our simulations (see Table 1).

\subsection{Spatial Distributions of Shock Waves and Shock-Heated Gas}

In this subsection, we investigate the spatial distribution of
shock-heated gas in conjunction with that of shock waves themselves. As
mentioned in the introduction, two types of shocks can be distinguished.
``External shocks'' are accretion shocks around nonlinear structures,
whose speed is basically the infall velocity of accreting flows.
``Internal shocks'' are products of infall, merging, and chaotic
flow motions induced by gravitational interactions inside nonlinear
structures. So the shock speed of both types should reflect the depth
of gravitational potential wells of nonlinear structures, and hence
their morphology. In our simulations, 1) high-speed shocks with
$v_{sh} \ga 700 \kms$ are found mostly around and inside knot-like
structures of clusters/groups, 2) shocks with
$150 \kms \la v_{sh} \la 700 \kms$ mostly around and inside filamentary
structures, and 3) low-speed shocks with $v_{sh} \la 150 \kms$ mostly
around and inside sheet-like structures. The left panels of Figure 2
show the locations of external shocks, divided into the three speed
groups, in a two-dimensional slice at $z=0$ from the $1024^3$ simulation.
The shocks encompass and outline nonlinear structures. The structures
with width of $\sim 1 h^{-1}{\rm Mpc}$ in the top-left panel are mostly
sheet-like structures with shallow potentials, while the thicker structures
with width of $\ga 2 h^{-1}{\rm Mpc}$ in the mid-left panel are filaments
(see also Fig. 3 below). The structure in the bottom-left panel contains
a cluster with X-ray emission-weighted temperature, $T_x\approx 2.4$ keV,
and X-ray luminosity, $L_x\approx 4.1\times10^{45}h$ erg s$^{-1}$ 
(note this cluster has a high luminosity for its temperature due to
excessive cooling without feedbacks from stars and galaxies).

The morphological distinction can be revealed more clearly in
three-dimensional volume renderings for the distribution of shocks waves.
The image in Figure 3(a) displays all shocks (external and internal)
inside the full box of (100 $h^{-1}{\rm Mpc}$)$^3$. At a quick glance,
filamentary structures look dominant. However, a careful inspection,
especially near the edges of the box where structures do not overlap,
exhibits lots of of sheet-like structures. In order to take a close
look at sheet-like structures, we show in Figures 3(c) and (d)
the distribution of low speed shocks with $v_{sh} < 150 \kms$ in a region
of (25 $h^{-1}{\rm Mpc}$)$^3$, which is marked in the top-left panel of
Figure 2. For filamentary structures we show in Figures 3(e) and (f)
the distribution of shocks with $150 \kms < v_{sh} < 700 \kms$ in a region
of (31 $h^{-1}{\rm Mpc}$)$^3$, which is marked in the mid-left panel of
Figure 2. For further reference, volume rendering movies for shocks of
different speed groups in the two regions shown in Figure 3(c) to (f)
as well as in the region covering the complex in the bottom-left panel
of Figure 2 are posted at ``http://canopus.chungnam.ac.kr/ryu/shock.html''.
The figures (along with the movies) demonstrate that the morphology of
nonlinear structures can be revealed through the distributions of shocks
of the three speed groups.

Naturally, it is expected that the spatial distribution of shock-heated
gas is closely related to that of cosmological shock waves. Figure 3(b)
shows a three-dimensional volume rendering image for the temperature of
shock-heated gas in the same perspective as that of Figure 3(a).
The images confirms that the two distributions are indeed very similar.
The right panels of Figure 2 show the gas distribution in three
temperature ranges, $10^3 < T < 10^5 \K$, $10^5 < T < 10^7 \K$,
and $T > 10^7 \K$, in the same two-dimensional slice as in the left
panels. With the choice of temperature ranges, different morphology
defined by shock waves is revealed once again by gas temperature too:
\ie the hot component mostly in knot-like structures, the WHIM with
$T=10^5-10^7 \K$ mostly in filaments, and the low temperature WHIM with
$T < 10^5\K$ mostly in sheet-like structures. Hence our results suggest
the existence of a component of the intergalactic gas, the low
temperature WHIM with $T < 10^5\K$, that has not been explored in details
in previous studies.

\section{Shock-Heated Gas in Observations }

\subsection{X-Ray Emission from Shock-Heated Gas}

As noted in the introduction, it was suggested that the X-ray
emission from the WHIM may contribute significantly to the soft XBR.
\citet{poc01} estimated that the emission would contribute
$\sim 10 \%$ of the observed XBR in the $0.5-2$ keV range. They
used $\Lambda$CDM simulation data in which radiative cooling,
feedbacks from galaxies and photo-ionization were included, and
$512^3$ grid zones were used for the box of $100 h^{-1}{\rm Mpc}$
size. On the other hand, based on analyses of observational data,
\citet{zmmg04a} suggested that most of the soft XBR could 
come from the WHIM. In order to asses how numerical
details including finite resolution would affect the prediction,
we calculated the X-ray emission from the shock-heated
gas outside clusters/groups using our simulation data with $1024^3$,
$512^3$, and $256^3$ grid zones. Similarly as \citet{poc01}, the gas
in spherical regions of $1 h^{-1}{\rm Mpc}$ radius around
clusters/groups with $T_x \geq 1$ keV was excluded, since they
would be identified as discrete X-ray sources and be removed from
contributors to the background radiation. Also the gas below
the Lyman-$\alpha$ and fast-cooling lines in Figure 1 was excluded.
The X-ray spectrum code for an optically thin gas, MEKAL \citep{mgo85},
was used to calculate the mean proper volume emissivity, $\epsilon(E,z)$,
from the entire simulation box at $z=2$, 1.5, 1, 0.5, 0.2, and 0.
The mean background intensity at $z=0$, $J(E)$, was calculated then
by integrating $\epsilon(E,z)$ from $z=2$ to $z=0$. Also the energy
flux, $dF/d\log T$, in the X-ray band of $0.5-2$ keV from the gas with
temperature between $\log T$ and $\log T + d\log T$ was calculated
from the proper volume emissivity, $\epsilon(E,T,z)$, as in the
calculation of $J(E)$. The upper and lower panels of Figure 4 show
$J(E)$ and $dF/d\log T$, respectively, from the simulation data
of different resolution.

Two points are noticed in Figure 4.
1) From the bottom panel, we confirm that among the shock-heated
gas, it is the WHIM with $T \sim$ $5 \times 10^6-10^7 \K$ that
contributes most to the soft XBR.
2) The amount of X-ray emission increases with resolution,
although the amount of shock-heated gas that emits such radiation
decreases (see Table 1). It is because there is systematically
more gas with higher density in higher resolution simulations.
The X-ray emission in our two high resolution simulation
($1024^3$ and $512^3$) differs by only a factor of two or so in
$E \ga 0.5$ keV. The fractional contribution to the observed XBR
\citep[see][for discussions on observations]{poc01} is largest in
the $0.5-1$ keV range, and estimated to be $\sim 30 \%$ in the
$1024^3$ simulation and $\sim 15 \%$ in the $512^3$ simulation.
Our estimation from the $512^3$ simulation is comparable to,
although a little larger than, $\sim 10 \%$ of \citet{poc01}, which
has the same spatial resolution. Although our simulations and that
in \citet{poc01} were performed with the same numerical code,
different treatments of additional physical processes as well as
details of numerical schemes should have led to the difference.
However, our estimation from the highest resolution simulation with
$1024^3$ grid zones is $\sim 3$ times larger than theirs, indicating
that the WHIM could be a major contributor to the soft XBR.

\subsection{Absorption Systems of Shock-Heated Gas}

It was suggested that the properties of the shock-heated gas in
filaments and sheets may be studied best by analyzing the soft X-ray
and far UV ``absorption systems'' or ``X-ray forests'' in the spectra
of distance quasars and AGNs. Especially, the OVI absorption systems
in far UV and the OVII and OVIII absorption systems in soft X-ray
were studied extensively by several authors (see \S 1 for references).
We also study the properties of such absorbers in our simulated local
universe at $z=0$, following the procedure described below. First,
we constructed a table for the fractions of Oxygen ions as a function
of gas temperature and density. Since photo-ionization of Oxygen ions
is important for the low density gas in filaments/sheets, the X-ray
background radiation of \citet{miou98} and the UV background radiation
of \citet{srgp99} were included, following \citet{cwkd03}. The
photo-ionization code, CLOUDY \citep{ferland98}, was used to compute
the fractions of Oxygen ions in a grid of temperature and density.
Figure 5 shows the resulting fractions of Oxygen ions at hydrogen number
density $-7 \leq \log n_H \leq -4$ as a function of temperature.
For comparison, the fractions when no ionizing background radiation
presents are also shown. It is obvious that the photo-ionization of
Oxygen ions by the adopted background radiations is very significant
in the temperature range that corresponds to filaments and sheets
($T < 10^7 \K$). The most noticeable point is that, with the ionizing
background radiations, the fractions of highly ionized Oxygen ions
increase at lower density. According to Figure 1 the WHIM
have the hydrogen number density, $-6 \la \log n_H \la -5$, so the
fractions of OVI, OVII and OVIII ions are highest in those components.
This is one of the reasons why these Oxygen ions were studied
extensively in previous works.

Next we calculated the column density distributions of absorption systems
due to these Oxygen ions. In simulation data, clouds of shock-heated gas
were identified along the $N_{lp} = 3 \times N_g^2$ ($N_g$ the number
of grid zones in one-direction) line paths of $L = 100 h^{-1}{\rm Mpc}$
at $z=0$. Again the gas in spherical regions of $1 h^{-1}{\rm Mpc}$ radius
around clusters/groups with $T_x \geq 1$ keV as well as the gas below 
the Lyman-$\alpha$ and fast-cooling lines were excluded. The metallicity
was assigned at each grid zone according to equation 
(\ref{zmetal}).
The left panels of Figure 6 show the cumulative distributions of
absorption system column density for OV-OVIII ions:
\begin{equation}
{dN(>N_{\rm O~ions}) \over dz} = {N_{\rm abs}(>N_{\rm O~ions}) \over
\Delta z},
\end{equation}
where $N_{\rm abs}$ is the number of absorption lines with
column density, $N_{\rm O~ions}$, greater than given values.
As expected from Figure 5, the OVII and OVIII absorption systems are
strongest. In the $1024^3$ simulation data, $dN/dz \sim 1$ for both
$N_{\rm OVII}$ and $N_{\rm OVIII} \ge 10^{15}{\rm cm^{-2}}$, which is
in a good agreement with \citet{cwkd03} and \citet{vbcm03}. The column
densities of OV and OVI ions are smaller, so $dN/dz \sim 1$ for
$N_{\rm OV}\ga 3\times 10^{14}{\rm cm^{-2}}$ and
$N_{\rm OVI} \ga 2\times 10^{14}{\rm cm^{-2}}$. We note the column
density distributions in the $1024^3$ and $512^3$ simulation data are
converged within a factor of two. 

The right panels of Figure 6 show the mass fractions of OV-OVIII ions
due to the gas with temperature between $\log T$ and $\log T + d\log T$,
\begin{equation}
{df_{\rm O~ions}(T) \over d\log T} = {1 \over {\cal M}_{\rm O}}
{d{\cal M}_{\rm O~ions}(T) \over d\log T},
\end{equation}
which was normalized with the total Oxygen mass ${\cal M}_{\rm O}$.
We find that the WHIM with $T=10^5-10^7 \K$ contributes to most of the
absorption systems of OVII ($\sim 70 \%$) and OVIII ($\sim 80 \%$).
Hence the observations of those absorption systems would explore
filamentary structures, as pointed out in \citet{vbcm03}. However,
the low temperature WHIM with $T < 10^5\K$ is the major contributor to
the OV and OVI absorption systems, accounting for $\sim 55 \%$ and
$\sim 60 \%$, respectively. This emphasizes the importance of the low
temperature WHIM in some observations, as noted by
\citet{bergeron02} in the study of OVI systems in a quasar spectrum.

\subsection{Emission Lines from Shock-Heated Gas}

Observations of emission lines from the shock-heated gas in filaments
and sheets still remain technically challenging. However, it was
suggested that detections of such emissions could be possible in future
X-ray missions such as MBE (see http://www.ssec.wisc.edu/baryons/)
and DIOS \citep{oisi04}. In this subsection, we study the Oxygen
emission lines radiated by the WHIM in our simulation data.
First, a table for the emissivity of OIII (698 \AA), OIV (549 \AA),
OV (630 \AA), OVI (1032 \AA), OVII (574 eV), and OVIII (653 eV) lines
was constructed on a grid of gas temperature and density, using CLOUDY
\citep{ferland98}. Collisional ionization as well as photo-ionization
due to the UV and X-ray background radiations described in \S 4.2 were
included. Figure 7 shows the resulting emissivity from the optically
thin gas of hydrogen number density $-7 \leq \log n_H \leq -4$. For
comparison, the emissivity when no ionizing background radiation presents
is also shown. From this figure one can expect that for $T > 10^6 \K$
the strongest lines from the gas with $-6 \la \log n_H \la -5$ in
filaments and sheets (see Fig. 1) would be the OVII and OVIII lines.
However, for $T < 10^6 \K$, OV and OVI would produce emission lines
stronger than OVII and OVIII lines.

Using the emissivity table, the Oxygen line emissivity from the gas
in simulation data was calculated at $z=0$. As in the calculation of
absorption column density, only the shock-heated gas was included and
the same metallicity-density relation was adopted. The left panels of
Figure 8 show the mean emissivity, $j_{\rm O~lines}$, of OV - OVIII
lines from the gas with temperature between $\log T$ and
$\log T + d\log T$. Overall the strength is greatest for the OV and OVI
lines, followed by the OVII and OVIII lines. While virtually all
emissions of the OVII and OVIII lines come from the higher temperature
WHIM with $10^6 < T < 10^7 \K$ ($\sim 95 \%$ for both lines),
the OV and OVI lines are emitted mostly from the WHIM
with $10^5 < T < 10^6 \K$, $\sim 75 \%$ and $\sim 80 \%$,
respectively. Contrary to the OV and OVI absorptions which get
significant contributions from the WHIM with $T < 10^5\K$, the line
emissions of OV (540 \AA) and OVI (1032 \AA) come mostly from the WHIM
with $T > 10^5\K$. Our estimates of line emissions are converged again
within a factor of two in the $1024^3$ and $512^3$ simulation data.

Our results indicate that the exploration of shock-heated gas through
emissions in soft X-ray, such as the MBE and DIOS missions, would pick
up a fraction of the WHIM with $10^6 \la T \la 10^7 \K$ ($\sim 10 \%$
of total gas mass in our estimate), missing the lower temperature WHIM
with $T \la 10^6\K$. A search mission using the OV and OVI line emissions
in far UV would detect a larger fraction of the low temperature WHIM with
$10^5 \la T \la 10^6 \K$ ($\sim 14 \%$ of total gas mass), complementing
the proposed missions in the soft X-ray band.

We also calculated the specific intensity of Oxygen lines,
$I_{\rm O~lines}$, by integrating the emissivity/$4 \pi$
along columns of $100 h^{-1}{\rm Mpc}$ at $z=0$. Note that the Hubble
expansion speed over the scale of $100 h^{-1}{\rm Mpc}$ is $(1/30)c$,
and the angular size of a single grid of $\Delta l$ at a distance
of $L = 100 h^{-1}{\rm Mpc}$ is $\Delta l/L$ radian. So $I_{\rm O~lines}$
corresponds effectively to the intensity measured over a band of width
$\Delta\lambda / \lambda \sim 1/30$ with a beam of solid angle
$3.4' \times 3.4'$ or smaller, when computed with the $1024^3$ simulation
data. We integrated along $N_{lp} = 3 \times N_g^2$ different line paths
in the simulation box. The right panels of Figure 8 show the cumulative
fraction of line paths with $I_{\rm O~lines}$ larger than given values.
They show that the fraction of randomly-chosen line paths with
$I_{\rm O~lines} \ge 10^{-9}$ erg cm$^{-2}$ s$^{-1}$ sr$^{-1}$ would be
$\sim 10^{-2}$ for the OV and OVI lines, while it would be
$2 - 3 \times 10^{-3}$ for the OVII and OVIII lines. However, the fractions
of $I_{\rm O~lines} \ge 10^{-8}$ erg cm$^{-2}$ s$^{-1}$ sr$^{-1}$ are 
$f_N \sim 1 - 2 \times 10^{-4}$ for the OVII and OVIII lines and larger
than those for the OV and OVI lines. This reflects the fact that while
the OV and OVI lines are stronger overall, the OVII and OVIII lines are
from the WHIM with $T \ga 10^6 \K$ and density higher than the mean
density of shock-heated gas.

\section{Summary}

Hierarchical clustering induces cosmological shock waves in
the course of large scale structure formation in the universe
\citep{rkhj03}. The intergalactic gas was heated mostly by such
shocks \citep{co99a,dcob01,vss02}. In this paper, we studied
the properties, spatial distribution, and possible observational
manifestations of the shock-heated gas in filaments and sheets
with a set of simulation data of a $\Lambda$CDM universe from
a grid-based N-body/hydrodynamic code \citep{rokc93}.
The nature of fixed grid-spacing makes such data suitable to the
study of the shock-heated gas in filaments and sheets. Shock-heating
was focused, so feedbacks from stars, galaxies and AGNs, and
photo-ionization heating were ignored, although the photo-ionization
of metal ions by the background radiation were taken into account
in post-processing data analyses. We expect that exclusion of such
processes should not have weakened significantly the main results of
this work, since the gas in filaments and sheets was heated mostly
by cosmological shocks.

The speed of cosmological shock waves reflects the depth of gravitational
potential of the associated nonlinear structures, so their morphology.
We saw that 1) the shocks with $v_{sh} \ga 700 \kms$ are distributed
mostly around and inside clusters/groups, 2) the shocks with
$150\kms \la v_{sh} \la 700 \kms$ are mostly around and inside
filamentary structures, and 3) the shocks with $v_{sh} \la 150 \kms$
are mostly around and inside sheet-like structures. The distribution
of the shock-heated gas should be closely related to that of shock waves.
We found that the WHIM with $T=10^5-10^7 \K$ is distributed mostly
in filamentary structures, while the low temperature WHIM with $T < 10^5$K
is mostly in sheet-like structures. The hot gas with $T > 10^7$K resides
on knot-like structures as the intra-cluster/intra-group medium and
the low density medium around them. The amount of shock-heated gas was
estimated as follows: the hot component with $T > 10^7 \K$ (excluding
the gas inside clusters/groups) accounts for $\sim 2.2\%$ of total gas
mass, the WHIM with $T=10^5-10^7 \K$ for $\sim 24\%$, and the WHIM with
$T < 10^5 \K$ for $\sim 13\%$ in our highest resolution simulation.
Thus our results indicate the existence of the WHIM with $T < 10^5 \K$,
which was heated by shocks of low speed with $v_{sh} \la 150 \kms$ and
is distributed as sheet-like structures. We suggest that {\it the low
temperature WHIM with $T < 10^5 \K$ contributes to a significant
fraction of the missing baryons at low redshifts} \citep{fhp98,fuk03}. 

It has been suggested in previous studies (see \S 1 for references)
that the shock-heated gas in filaments and sheets can be manifested
through emissions and absorptions in soft X-ray and far UV.
1) We found that the soft X-ray emission from the shock-heated gas 
in our highest resolution data contributes to $\sim 30 \%$ of the
observed XBR in the $0.5-1$ keV range. This is somewhat larger than that
previously suggested by \citet{poc01}.
2) Column densities of Oxygen ions along randomly selected line paths 
in the computational box were calculated. Such column densities can
produce the absorption systems or X-ray forests in the spectra of distant
quasars and AGNs \citep[see, \eg][]{hgm98}. The photo-ionization of Oxygen
ions by the background UV and X-ray radiations is important for the gas
in filaments and sheets due to low density and temperature. As a result,
the fractions of Oxygen ions depend strongly on the local gas density.
We found that, for OVII and OVIII, the absorption systems associated
with the WHIM with $T=10^5-10^7 \K$ account for $70 - 80 \%$ of the
identified absorption systems, while the low temperature WHIM with
$T < 10^5\K$ contributes $\sim 55 - 60 \%$ of the OVI and OV absorption
systems.
3) Finally, the emission lines from Oxygen ions were calculated. They
are emitted mostly from collisionally excited ions. We estimated that
$\sim 95 \%$ of the OVII line (574 eV) and OVIII line (653 eV) emissions
come from the WHIM with $10^6 < T < 10^7 \K$, while $\sim 75 \%$ of
the OVI line (1032 \AA) and $\sim 80 \%$ of the OV line (630 \AA) are
emitted by the WHIM with $10^5 < T < 10^6 \K$. Hence, we conclude that
{\it absorption systems in UV due to species such as OV and OVI provide
the best chance to detect the low temperature WHIM with $T < 10^5\K$.}

\acknowledgements
HK and DR were supported in part by KOSEF through Astrophysical
Research Center for the Structure and Evolution of Cosmos (ARCSEC)
and grant R01-1999-00023. HK and DS were supported in part by a grant
at Korea Astronomical Observatory. Numerical simulations utilized
``The Grand Challenge Program'' of the KISTI Supercomputing Center.
We thank T. W. Jones, E. Hallman and R. Benjamin for discussions,
and anonymous referee for constructive comments.

\clearpage

\begin{deluxetable} {cccc}
\tablecaption{Mass fractions of shock-heated gas}
\tablehead{
\colhead {Component} & \colhead{$1024^3$} & \colhead{$512^3$}
& \colhead{$256^3$}
}
\startdata
Hot with $T > 10^7 \K$\tablenotemark{a} & 0.022  & 0.024  & 0.029 \\
WHIM with $T=10^5-10^7 \K$              & 0.24 &   0.37 &   0.43 \\
WHIM with $T < 10^5 \K$                 & 0.13 &   0.13 &   0.10 \\
\enddata 
\tablenotetext{a}{The gas within 1 $h^{-1}$Mpc of clusters/groups hotter
than 1keV was not counted.}
\end{deluxetable}

\clearpage

\begin{figure}
\figcaption{Mass distribution of shock-heated gas in the plane of gas
density and temperature at $z=0$ in simulations with different
resolution. For comparison, the distribution from the $1024^3$ simulation
without cooling is also shown. Purple indicates the highest mass
concentration, and the color scale was set arbitrary to highlight
the distribution. The black straight line presents the ``Lyman-$\alpha$
equation of state'' and the black curved line isolates the ``fast-cooling''
region of \citet{vss02} (see text for details).}
\end{figure}


\begin{figure}
\figcaption{{\it Left panels}: Locations of external shocks at
$z=0$ in a two-dimensional slice of (100 $h^{-1}{\rm Mpc}$)$^2$
from the $1024^3$ simulation. The shocks are classified into
three groups according to shock speed, $v_{\rm sh} < 150 \kms$,
$150 < v_{\rm sh} < 700 \kms$, and $ v_{\rm sh} > 700 \kms$,
respectively, from top to bottom.
{\it Right panels}: Spatial distribution of shock-heated gas
at $z=0$ in the same slice in three ranges of temperature:
the WHIM with $T < 10^5 \K$, the WHIM with $T=10^5-10^7 \K$,
and the hot component with $T > 10^7 \K$, shown in the top,
middle and bottom panels, respectively. In each panel purple
represents to the gas with temperature close to the upper bound
and green close to the lower bound, while blue in the middle.}
\end{figure}


\begin{figure}
\figcaption{(a) Distribution of shock waves (external and internal) at
$z=0$ in the full box of (100 $h^{-1}{\rm Mpc}$)$^3$. Color represents
logarithmically-scaled shock speed from $v_{\rm sh} = 15 \kms$ (yellow)
to $1,500 \kms$ (purple) and higher.
(b) Temperature distribution of shock-heated gas at $z=0$ in the same
box. Color covers logarithmically-scaled temperature from $T = 10^4 \K$
(yellow) to $10^8 \K$ (purple) and higher.
(c, d) Distribution of shock waves (external and internal) with
$v_{\rm sh} = 15 \kms$ (yellow) to $150 \kms$ (blue-green) at $z=0$
in a region of (25 $h^{-1}{\rm Mpc}$)$^3$, marked at the top-left
panel of Figure 2. Tow different perspectives are shown.
(e, f) Distribution of shock waves (external and internal) with
$v_{\rm sh} = 150 \kms$ (green) to $700 \kms$ (red) at $z=0$
in a region of (31 $h^{-1}{\rm Mpc}$)$^3$, marked at the mid-left
panel of Figure 2. Tow different perspectives are shown.
All images are from the $1024^3$ simulation.}
\end{figure}

\begin{figure}
\vspace{-1.5cm}
\hspace{-1.5cm}
\epsfxsize=18cm\epsfbox{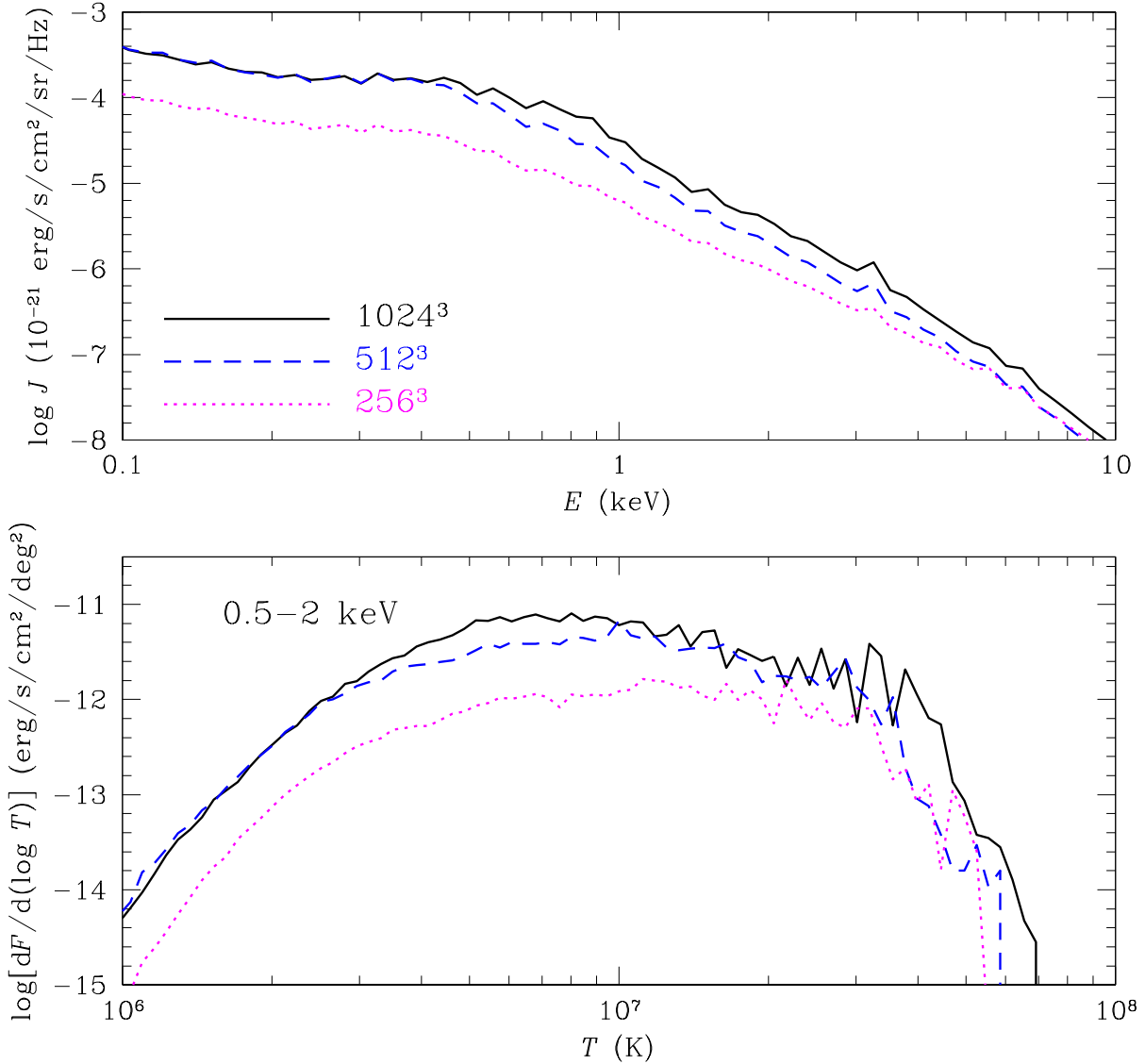}
\vspace{-2cm}
\figcaption{{\it Upper panel}: Mean background intensity in soft X-ray
from shock-heated gas, integrated from $z=2$ to $z=0$, in simulations
with different resolution.
{\it Lower panel}: X-ray flux in the $0.5-2$ keV energy band
from shock-heated gas as a function of temperature, integrated
from $z=2$ to $z=0$, in the same simulations. The gas in spheres of
a radius 1 $h^{-1}$Mpc around clusters/groups hotter than 1 keV
was excluded in calculating the X-ray radiation.}
\end{figure}

\begin{figure}
\vspace{-0.5cm}
\hspace{-1.3cm}
\epsfxsize=18cm\epsfbox{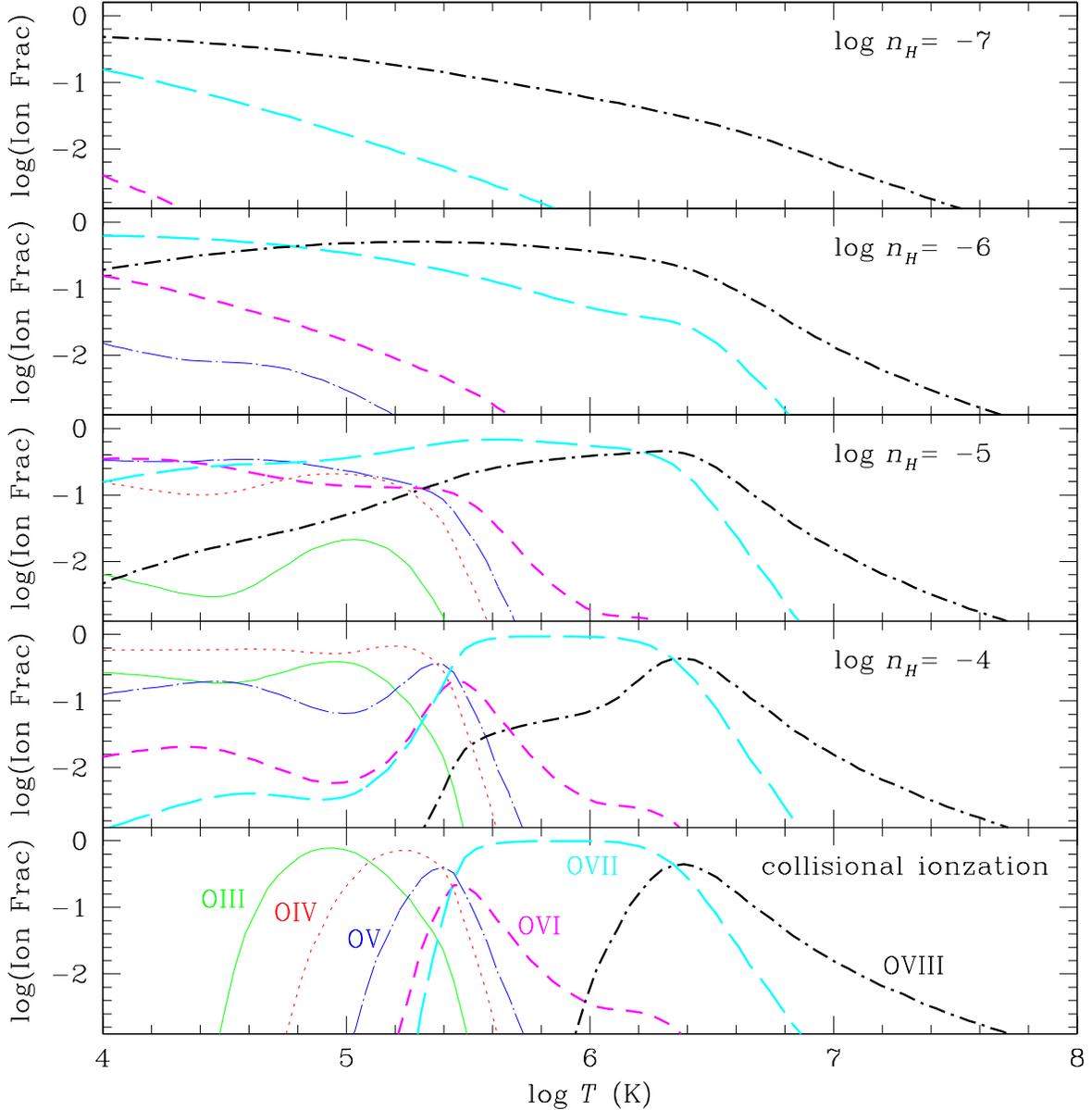}
\vspace{-1.3cm}
\figcaption{Fractions of OIII to OVIII ions as a function of
temperature for the gas of hydrogen number density $\log(n_H)= -7$
to $-4$. Collisional ionization as well as photo-ionization by
the UV and X-ray background radiations \citep{miou98,srgp99}
were included. For comparison, the fractions of the ions in
collisional ionization equilibrium are also shown.}
\end{figure}

\begin{figure}
\vspace{-0.5cm}
\hspace{1cm}
\epsfxsize=14cm\epsfbox{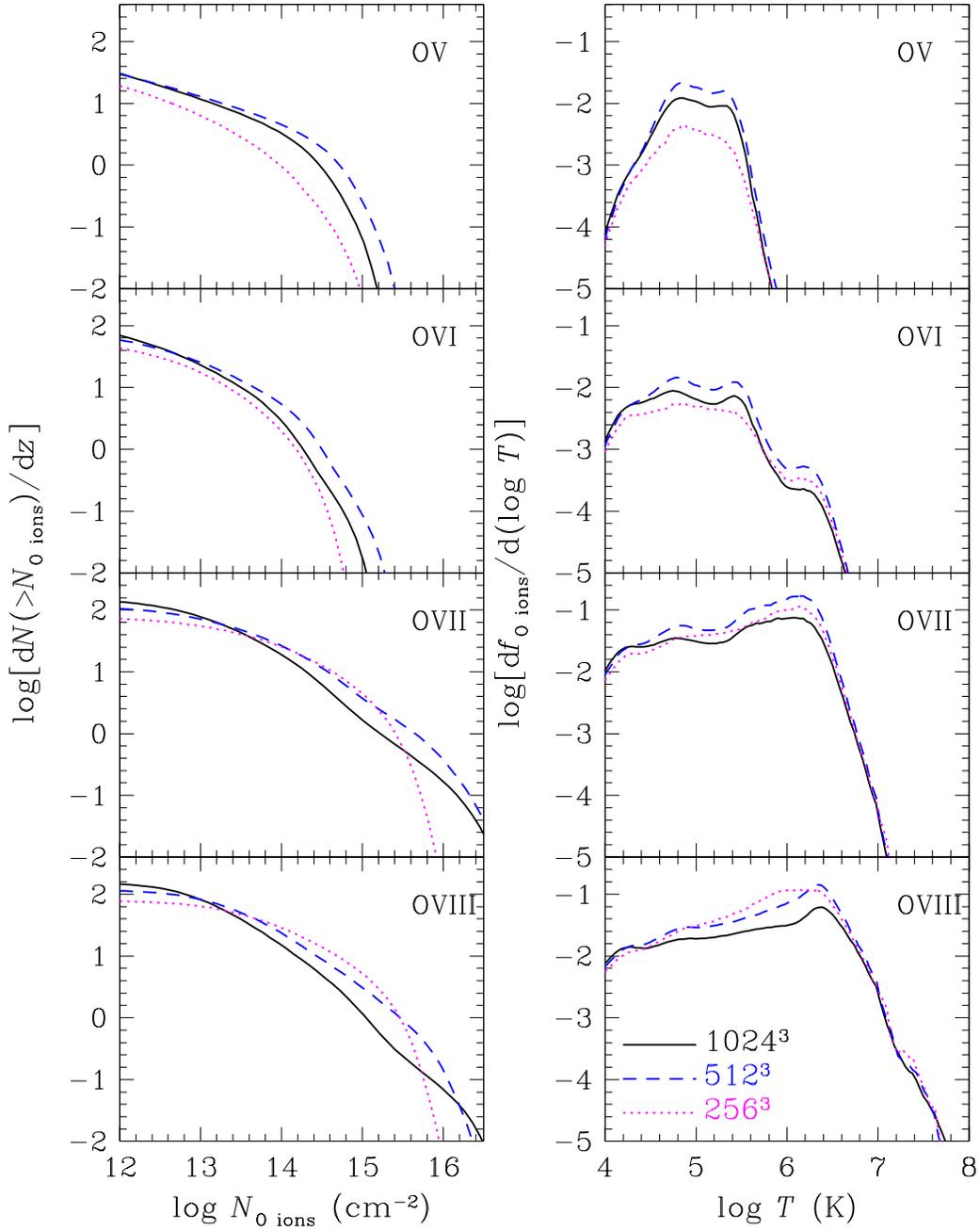}
\vspace{-1cm}
\figcaption{{\it Left panels}: Number of absorption lines per unit
redshift by intervening gas clouds at $z=0$ with the column density of
Oxygen ions larger than $N_{\rm O~ions}$. 
{\it Right panels}: Fractions of OV - OVIII ions relative to the 
total Oxygen abundance in shock-heated gas at $z=0$ as a function of
temperature.} 
\end{figure}

\begin{figure}
\vspace{-0.5cm}
\hspace{-1.3cm}
\epsfxsize=18cm\epsfbox{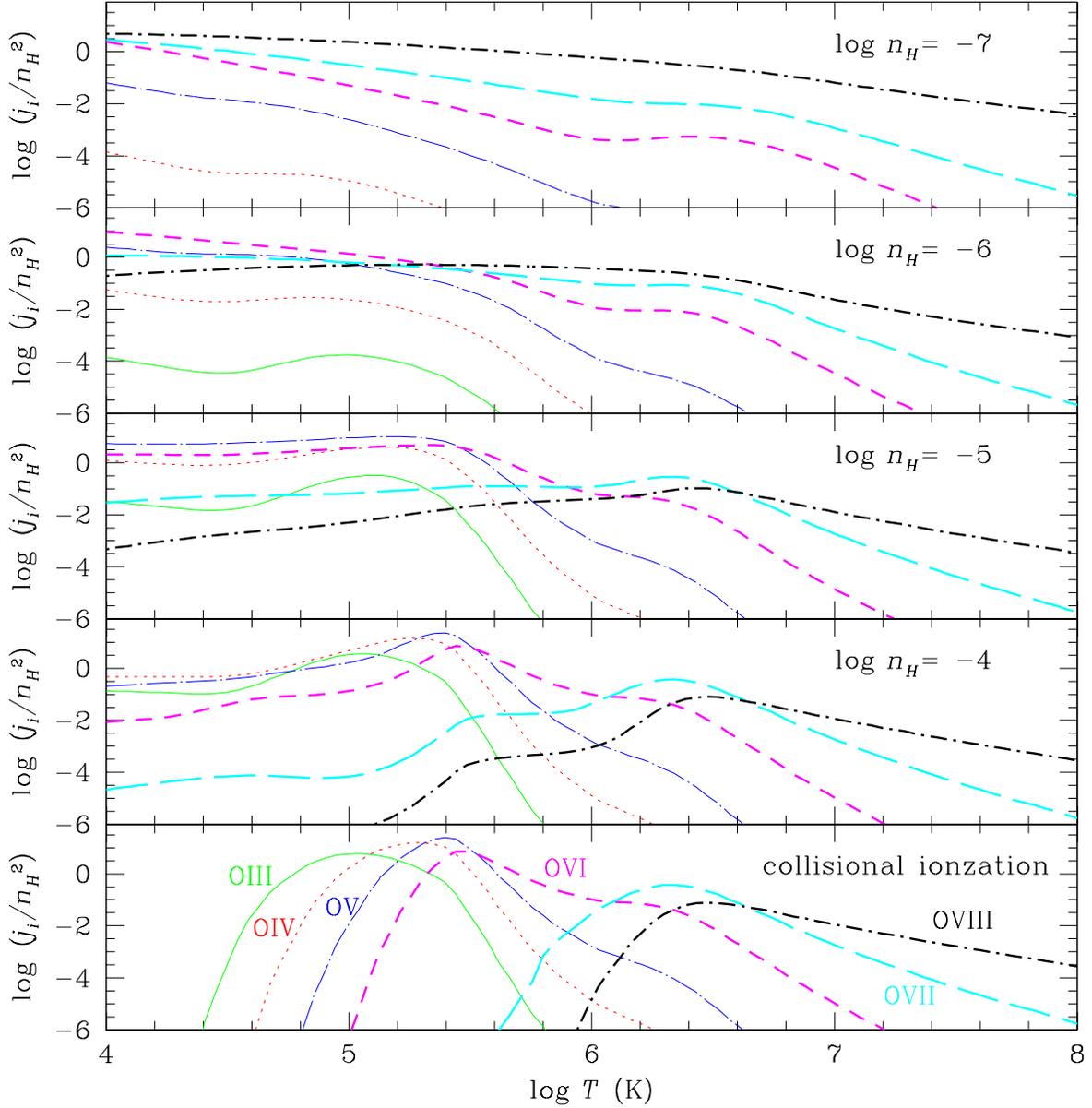}
\vspace{-1.3cm}
\figcaption{Emissivity of OIII to OVIII ion lines as a function of
temperature for the gas of hydrogen number density $\log(n_H)= -7$ to
$-4$: OIII (698 \AA), OIV (549 \AA), OV (630 \AA), OVI (1032 \AA),
OVII (574 eV), OVIII (653 eV). $j_i/n_H^2$ is in units of $10^{-24}
{\rm erg~cm^3~s^{-1}}$. Collisional ionization as well as
photo-ionization by the UV and X-ray background radiations
\citep{miou98,srgp99} were considered. For metallicity, $Z=0.1Z_{\sun}$
was set. For comparison, the emissivity of the ion lines in collisional
ionization equilibrium is also shown.}
\end{figure}

\begin{figure}
\vspace{-0.5cm}
\hspace{1cm}
\epsfxsize=14cm\epsfbox{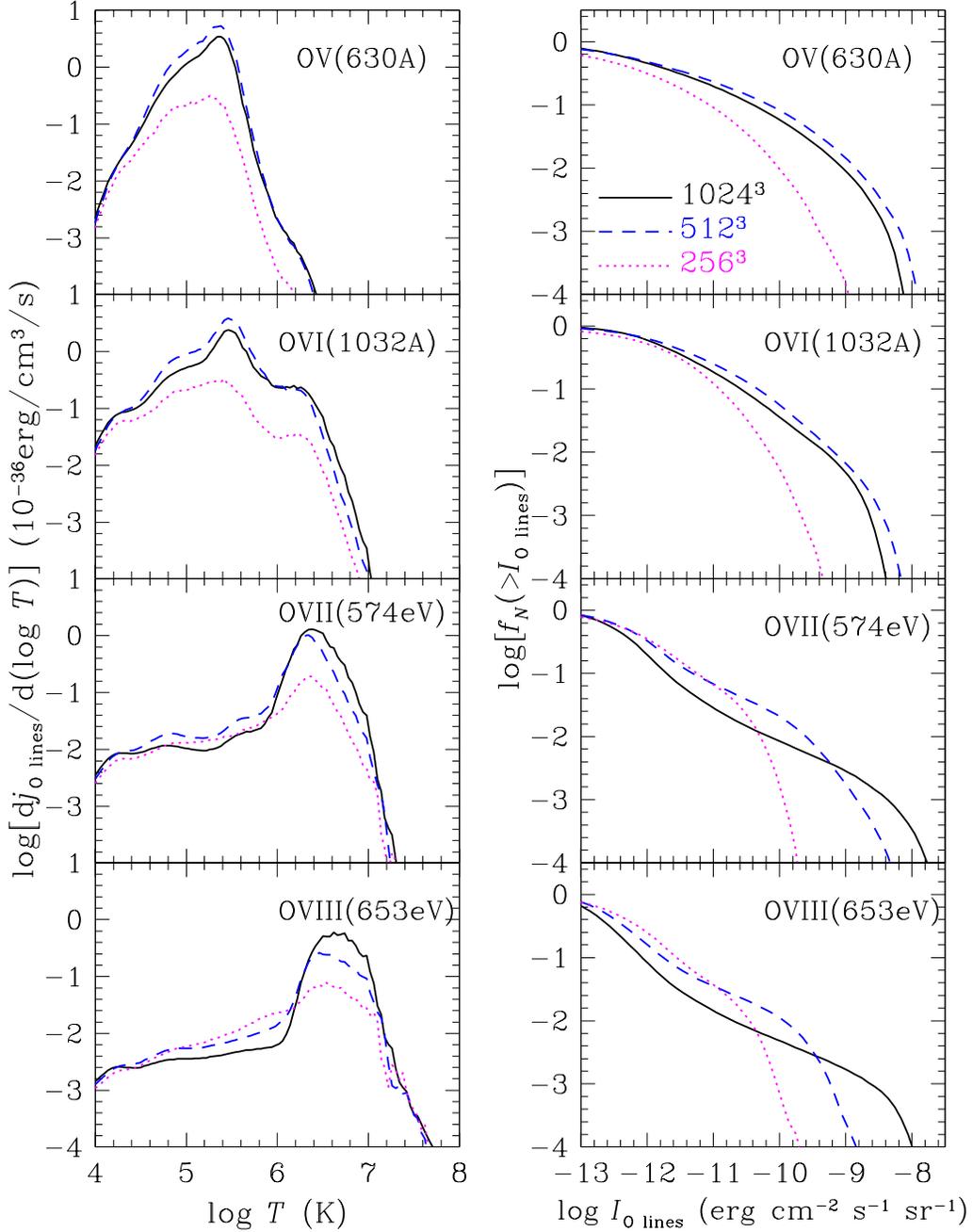}
\vspace{-1cm}
\figcaption{{\it Left panels}: Mean emissivity of OV to OVIII lines
from shock-heated gas at $z=0$ as a function of temperature.
{\it Right panels}: Cumulative fraction of line paths with the
specific intensity of Oxygen lines larger than $I_{\rm O~lines}$.
The intensity was calculated by integrating the emissivity divided
by $4\pi$ along a length of $100 h^{-1}{\rm Mpc}$ at $z=0$.}
\end{figure}


\begin{thebibliography}{}


\bibitem[Bennett \etal(2003)]{bhhj03}
Bennett, C. L. \etal~
2003, \apjs, 148, 1

\bibitem[Bergeron \etal(2002)]{bergeron02}
Bergeron, J., Aracil, B., Petitjean, P, and Pichon, C.
2002, \aap, 396, L11

\bibitem[Cen \etal(1995)]{ckor95}
Cen, R., Kang, H., Ostriker, J. P., \& Ryu, D.
1995, \apj, 451, 436

\bibitem[Cen \& Ostriker(1999a)]{co99a}
Cen, R. \& Ostriker, J. P.
1999, \apj, 514, 1

\bibitem[Cen \& Ostriker(1999b)]{co99b}
Cen, R. \& Ostriker, J. P.
1999, \apj, 519, L109 

\bibitem[Cen \etal(2001)]{cen01}
Cen, R., Tripp, T. D., Ostriker, J. P. \& Jenkins, E. B., 2
2001, \apj, 559, L5 

\bibitem[Chen \etal(2003)]{cwkd03}
Chen, X., Weinberg, D. H., Katz, N., \& Dav\'e R.
2003, \apj, 594, 42

\bibitem[Dav\'e \etal(2001)]{dcob01}
Dav\'e, R. \etal~
2001, \apj, 552, 473


\bibitem[Fang \& Bryan(2001)]{fb01}
Fang, T., \& Bryan, G. L.,
2001, ApJ, 561, L31.


\bibitem[Fang \etal(2003)]{fcsh03}
Fang, T., \etal~
2003, preprint (astro-ph/0311141)


\bibitem[Ferland \etal(1998)]{ferland98}
Ferland, G. J., Korista, K. T., Verner, D. A., Ferguson, J. B., 
Kingdon, J. B., \& Verner, E. M.
1998, \pasp, 110, 761

\bibitem[Finoguenov(2003)]{fbh03}
Finoguenov, A., Briel, U. G., \& Henry, J. P.
2003, \aap, 410, 777

\bibitem[Frenk \etal(1999)]{fwbb99}
Frenk, C. S. \etal~
1999, \apj, 525, 554

\bibitem[Fukugita \etal(1998)]{fhp98}
Fukugita, M., Hogan, C. J., \& Peebles, P. J. E.
1998, \apj, 503, 518

\bibitem[Fukugita(2003)]{fuk03}
Fukugita, M.
2003, preprint (astro-ph/0312517)

\bibitem[Furlanetto \& Loeb(2003)]{fl03}
Furlanetto, S. R. \& Loeb, A.
2003, preprint (astro-ph/0312435)

\bibitem[Furlanetto \etal(2003)]{fssh03}
Furlanetto, S. R., Schaye, J., Springel, V., \& Hernquist, L.,
2003, preprint (astro-ph/0311006)

\bibitem[Gabici \& Pasquale(2003)]{gp03}
Gabici, S. \& Pasquale, B.
2003, \apj, 583, 695


\bibitem[Hellsten \etal(1998)]{hgm98}
Hellsten, U., Gnedin, N. Y., \& Miralda-Escud\'e, J.
1998, \apj, 509, 56

\bibitem[Hui \& Gnedin(1997)]{hg97}
Hui, L. \& Gnedin, N. Y.
1997, \mnras, 292, 27

\bibitem[Kang \etal(1994)]{kcor94}
Kang, H., Cen R., Ostriker, J. P., \& Ryu, D.
1994, \apj, 428, 1

\bibitem[Kang \etal(1996)]{krj96}
Kang, H., Ryu, D., \& Jones, T. W.
1996, \apj, 456, 422

\bibitem[Kang \etal(1994)]{kocr94}
Kang, H., Ostriker, J. P., Cen, R., Ryu, D., Hernquist, L., Evrard, A. E.,
Bryan, G. L., \& Norman, M. L.
1994, \apj, 430, 83

\bibitem[LeVeque(1997)]{leve97}
LeVeque, R. J.
1997, in 27th Saas-Fee Advanced Course Lecture Notes ``Computational Methods
in Astrophysical Fluid Flows'', ed. O. Steiner \& A. Gautschy
(Berlin: Springer)


\bibitem[Mewe \etal(1985)]{mgo85}
Mewe, R., Gronenschild, E. H. B. M., \& van den Oord, G. H. J.
1985, \aaps, 62, 197

\bibitem[McKernan \etal(2003)]{mymg03}
McKernan, B., Yaqoob, T., Mushotzky, R., George, I. M., \& Turner, T. J.
2003, preprint (astro-ph/0310476)

\bibitem[Miyaji \etal(1998)]{miou98}
Miyaji, T., Ishisaki, Y., Ogasaka, Y., Ueda, Y., Freyberg, M. J., 
Hasinger, G., \& Tanaka, Y.,
1998, \aap, 334, L13

\bibitem[Miniati \etal(2000)]{mrk00}
Miniati, F., Ryu, D., Kang, H., Jones, T. W., Cen, R., \& Ostriker, J.
2000, \apj, 542, 608

\bibitem[Nicastro \etal(2003)]{nica03}
Nicastro, F. \etal~
2003, \nat, 421, 719

\bibitem[Phillips \etal(2001)]{poc01}
Phillips, L. A., Ostriker, J. P., \& Cen, R.
2001, \apj, 554, L9

\bibitem[Quilis \etal(1998)]{qis98}
Quilis, V., Ibanez, J. M. A., \& Saez, D.
1998, \apj, 502, 518

\bibitem[Ryu \etal(2003)]{rkhj03}
Ryu, D., Kang, H., Hallman, E., \& Jones, T.W.,
2003, \apj, 593, 599

\bibitem[Ryu \etal(1993)]{rokc93}
Ryu, D., Ostriker, J. P., Kang, H., \& Cen, R.
1993, \apj, 414, 1

\bibitem[Shull \etal(1999)]{srgp99}
Shull, J. M., Robers, D., Giroux, M. L., Penton, S. V., \& Fardal, M. A.
1999, \aj, 118, 1450

\bibitem[Spergel \etal(2003)]{svpk03}
Spergel, D. N. \etal~
2003, \apjs, 148, 175

\bibitem[Sutherland \& Dopita(1993)]{sd93}
Sutherland, R., S. \& Dopita, M., A.
1993, \apjs, 88, 253

\bibitem[Sutherland \& Dopita(2002)]{sd02}
Sutherland, R., S. \& Dopita, M., A.
2002, in Astrophysics of the Diffuse Universe (Springer)

\bibitem[Ohashi \etal(2004)]{oisi04}
Ohashi, T. \etal, 2004, preprint (astro-ph/0402546)

\bibitem[Valageas \etal(2002)]{vss02}
Valageas, P., Schaeffer, R., \& Silk, J.
2002, \aap, 388, 741

\bibitem[Viel \etal(2003)]{vbcm03}
Viel, M., Branchini, E., Cen, R., Matarrese, S., Mazzotta, P.,
\& Ostriker, J. P.
2003, \mnras, 341, 792

\bibitem[Yoshikawa \etal(2003)]{yyso03}
Yoshikawa, K., Yamasaki, N. Y., Suto, Y., Ohashi, T., Mitsuda, K.,
Tawara, Y., \& Furuzawa, A.
2003, \pasj, 55, 879

\bibitem[Zappacosta \etal(2004a)]{zmmg04a}
Zappacosta, L., Maiolino, R., Mannucci, F., Gilli, R., Finoguenov, A.,
\& Ferrara, A.
2004a, preprint (astro-ph/0401202)

\bibitem[Zappacosta \etal(2004b)]{zmmg04b}
Zappacosta, L., Maiolino, R., Mannucci, F., Gilli, R., \& Schuecker, P.
2004b, preprint (astro-ph/0402575)

\end{thebibliography}
\end{document}